\documentclass[10pt,conference]{IEEEtran}

\usepackage{cite}
\usepackage{amsmath,amssymb,amsfonts}
\usepackage{algorithmic}
\usepackage{graphicx}
\usepackage{textcomp}
\usepackage{xcolor}
\usepackage{braket}
\usepackage{subcaption}
\usepackage{hyperref}
\usepackage{multirow}

\hypersetup{%
  colorlinks=true,%
  linkcolor={red!50!black},
  citecolor={blue!65!black},
  urlcolor={blue!80!black},
  bookmarksnumbered=true,%
  bookmarksopen=true}

\def\BibTeX{{\rm B\kern-.05em{\sc i\kern-.025em b}\kern-.08em
    T\kern-.1667em\lower.7ex\hbox{E}\kern-.125emX}}

\usepackage{tikz}
\usetikzlibrary{decorations.pathreplacing}
\usetikzlibrary{shapes.misc}
\usetikzlibrary{calc}
\usetikzlibrary{positioning}

\usepackage[inline]{enumitem}

\usepackage{lipsum}
\usepackage{cleveref}

\usepackage[nolist,printonlyused]{acronym}
\acrodef{QKD}{Quantum Key Distribution}
\acrodef{BQC}{Blind Quantum Computation}
\acrodef{CPS}{Classical Processing Unit}
\acrodef{QPS}{Quantum Processing Unit}
\acrodef{EDF}{Earliest Deadline First}
\acrodef{DQC}{Distributed Quantum Computing}

\newlist{inlinelist}{enumerate*}{1}
\setlist[inlinelist]{label=(\arabic*)}

\begin{document}

\title{Compilation strategies for quantum network programs using Qoala}

\author{
\IEEEauthorblockN{Samuel Oslovich}
\IEEEauthorblockA{\textit{QuTech, Delft University of Technology} \\
\textit{and Kavli Institute of Nanoscience} \\ Delft, The Netherlands \\ s.o.oslovich@tudelft.nl}
\and
\IEEEauthorblockN{Bart van der Vecht}
\IEEEauthorblockA{\textit{QuTech, Delft University of Technology} \\ \textit{and Kavli Institute of Nanoscience} \\ Delft, The Netherlands \\ b.vandervecht@tudelft.nl}
\and
\IEEEauthorblockN{Stephanie Wehner}
\IEEEauthorblockA{\textit{QuTech, Delft University of Technology} \\
\textit{and Kavli Institute of Nanoscience} \\
Delft, The Netherlands \\ s.d.c.wehner@tudelft.nl} 
}

\maketitle

\begin{abstract}
Execution of quantum network applications requires a software stack for nodes.
Recently, the first designs and demonstrations have been proposed for such software stacks, including QNodeOS and its extension, Qoala. The latter enables compilation strategies previously not possible in QNodeOS. Here, we show how the extensions provided by Qoala can be used by a compiler to improve the performance of quantum network applications. We define new compilation strategies that allow the compiler to influence the scheduling and execution of quantum programs on a quantum network node. Through simulation, we demonstrate that our compilation strategies can reduce the execution time by up to 29.53\% and increase the success probability by up to 25.12\%. Our work highlights the potential of compiler optimizations for quantum network programs.
\end{abstract}


\section{Introduction}
\label{sec:introduction}
Quantum networks consist of multiple connected quantum processors that use entanglement as a means to transfer quantum information and to realize correlations that are not possible classically.
Such networks enable applications~\cite{wehner_2018_stages} not possible on classical networks, including \ac{QKD}~\cite{bb84Original}, \ac{BQC}~\cite{broadbent_2009_ubqc,childs_2005_secure_qc}, data consistency in the cloud~\cite{benor_2005_byzantine}, privacy-enhancing proofs of deletion~\cite{poremba_quantum_2022}, and exponential savings in communication~\cite{guerin_exponential_2016}. 

We envision a future \emph{quantum internet} globally connecting quantum processors, in which independent nodes run their own programs that communicate with other nodes, realizing multi-program applications like those mentioned above. 
We consider applications all the way up to the most advanced stage~\cite{wehner_2018_stages}, including \emph{interactive} \emph{hybrid quantum-classical} applications~\cite{vandervecht2025qoala}. This means that programs consist of a hybrid of classical and quantum code segments, and that these segments may run concurrently, while influencing each other.
For example, a quantum code segment may `pause' halfway, keeping quantum states in memory, while waiting for a classical code segment to receive and process a message from another node.
This message could then determine further execution of the quantum code.

To realize quantum network applications in a scalable way, a set of software tools and abstractions is needed.
First, arbitrary programs need to be written in a high-level (quantum-) hardware-agnostic programming language, allowing developers to focus on behavior rather than physical details.
Second, a compiler must transform program code into efficient executable code.
Finally, a runtime environment on a node must execute the compiled code.
These constructs allow the development and execution of arbitrary quantum network applications, including applications that are still to be invented.

Recently, advancements have been made regarding architectures that can execute arbitrary quantum network applications on quantum processors.
QNodeOS was the first such full-stack system architecture, and it was demonstrated on real quantum hardware that it could execute applications programmed in a high level \cite{delle2025operating}.
Recently, QNodeOS was extended to improve compilation and scheduling, in an extension called Qoala \cite{vandervecht2025qoala}.

\subsection{Contributions}
We validate that hybrid quantum-classical compilation for network programs can provide performance improvements compared to QNodeOS style program execution (without hybrid compilation). Hybrid compilation refers to compilation applied on the joint quantum-classical program code, rather than on the quantum code and classical code separately. For our example program,  hybrid compilation can increase the success probability by up to 25.12\%. When considering BQC applications executed on near-term hardware, we identify limiting factors that reduce the improvements of hybrid compilation.
We identify new compilation strategies---\emph{block compilation}, \emph{deadline compilation}, and \emph{critical section compilation}---and show that they allow the compiler to influence how programs are executed, leading to improved program performance. We discover scenarios in which block compilation can decrease execution time by up to 29.53\%. We evaluate deadline compilation and identify its shortcomings. We identify conditions where critical section compilation can result in a decrease in execution time of up to 6.12\%. 

The remainder of this paper is structured as follows. \Cref{sec:related-works} explains how our work differs from the existing quantum compilation literature. \Cref{sec:background} provides context and revisits the Qoala architecture, highlighting the most relevant extensions to QNodeOS. In \cref{sec:sim-details} we briefly describe the Qoala simulator and relevant quantum hardware parameters. \Cref{sec:hybrid-comp} demonstrates performance improvements made possible through hybrid compilation. In sections \ref{sec:block-comp}, \ref{sec:deadline-comp}, and \ref{sec:critical-comp} we evaluate our new compilation strategies.

\section{Related Works}
\label{sec:related-works}
\subsection{Quantum Circuit Compilation}
Techniques for compiling quantum programs are well studied within the quantum computing community. Quantum compilation typically focuses on optimizing gate sequences, reducing circuit depth, and mapping logical qubits to physical architectures (see e.g. \cite{barenco_1995_Elementary_gates, maronese_2021_quantumcompiling, kusyk_2021_quantcircuitcompilationSurvey}). 
However, quantum network programs present unique challenges and opportunities for compilation. Quantum network programs often interleave classical and quantum control flow, resulting in a program that cannot be represented as a simple circuit. Additionally, quantum network applications consist of multiple programs that must be compiled on separate nodes. These programs consist of entanglement generation and classical communication, both of which are non-local operations that depend on other nodes and do not take a fixed amount of time to execute.

\subsection{Hybrid Quantum-Classical Compilation}
The general idea of hybrid quantum-classical compilation has previously been explored in the context of quantum computing \cite{Peduri_2022_QSSA, smith_2017_practicalquantuminstructionset, mintz_2019_qcorlanguageextensionspecification, bergholm_2022_pennylaneautomaticdifferentiationhybrid}. However, in quantum computing hybrid algorithms typically are not \emph{interactive}. That is, quantum states are not live in memory while classical operations are executed \cite{farhi_2014_quantumapproximateoptimizationalgorithm, McClean_2016_VQE}, or are restricted to low level, predictable interactions, like quantum error correction \cite{Lidar_2013_qerrcorrection} or mid-circuit measurements \cite{botelho_2022_midcircuitmeas}. On the other hand, quantum network programs often are \emph{interactive} (e.g.\cite{broadbent_2009_ubqc}), so quantum states are live in memory while waiting for input from another program. This results in more interesting and impactful compilation opportunities. Our work explores how hybrid quantum-classical compilation can be utilized to improve the performance of \emph{interactive} quantum network programs.

\subsection{Compilation for Distribution Quantum Computing}
Compilers for \ac{DQC} address the challenges of partitioning quantum circuits across multiple nodes in a network (see e.g. \cite{caleffi_2024_DQCSurvey, ferrari_2021_compilerdesignDQC, cuomo_2023_optimizedDQCcompiler}). These approaches consider the overhead and noise introduced by entanglement generation and distribution. However, \ac{DQC} compilation approaches have fundamentally different assumptions than those required for quantum network applications:
\begin{itemize}
    \item \textbf{Global Knowledge}: \ac{DQC} compilers generally assume complete knowledge of the entire quantum circuit to be executed and of the quantum hardware capabilities of each node, allowing for global optimization across all nodes \cite{caleffi_2024_DQCSurvey,ferrari_2021_compilerdesignDQC, cuomo_2023_optimizedDQCcompiler}. In contrast, quantum network scenarios often involve client-server applications where each node possesses only partial knowledge of the overall computation, requiring the individual programs be compiled separately at the client and server.
    \item \textbf{Limited Classical Processing}: \ac{DQC} typically executes a single distributed quantum circuits. These circuits are typically non-\emph{interactive}, with classical computation being limited to teleportation corrections. As previously mentioned, quantum network applications, may be \emph{interactive}, requiring input from another program while quantum states are live in memory. 
\end{itemize}

\subsection{Quantum Network Program Compilation}
Compiling arbitrary quantum network programs is a little explored research area. NetQASM \cite{dahlberg_2022_netqasm}, a quantum network assembly language, provides a compiler along with their software development kit, however no optimizations are implemented. To the best of our knowledge our work is the first to explore compilation strategies for improving the performance of arbitrary quantum network programs.

\section{Background}
\label{sec:background}

\subsection{Qoala recap}
\label{sec:background:qoala-recap}

We briefly summarize the Qoala architecture put forward in \cite{vandervecht2025qoala}.
This architecture can execute any arbitrary applications, such as \ac{QKD}, \ac{BQC}, or other (new) applications, as long as the application adheres to the Qoala format.
\emph{Applications} are realized as a collection (typically two or more) of single-node \emph{programs} that interact with each other.
A program runs on a single quantum network node, and its code is organized into \emph{blocks}.
Each block is of one of four types:
\begin{itemize}
    \item \emph{CL (classical local)}: classical computation. A block can contain arbitrary classical code; it is typically executed very fast. Using the classical processor from the QNodeOS experiment, it would take roughly 15ns per classical instruction \cite{delle2025operating, fog_2022_instructiontables}. See \Cref{sec:appendix:params} for more details.
    \item \emph{CC (classical communication)}: classical messaging with other programs (on other nodes). E.g. sending a quantum measurement outcome (classical bit). The time to receive a classical message depends on the distance and number of hops between nodes. In experiments in a lab setting, classical communication times have been measured at roughly 155$\mu$s \cite{delle2025operating,bovy_2002_internetdelay}.
    \item \emph{QL (quantum local)}: local quantum computation (such as gates and measurement). A QL block is expressed as a NetQASM subroutine~\cite{dahlberg_2022_netqasm} containing local quantum instructions. Executing quantum instructions takes much longer than classical instructions (roughly 50$\mu$s per quantum instruction in the QNodeOS experiment \cite{delle2025operating}).
    \item \emph{QC (quantum communication)}: entanglement generation with other programs (on other nodes). A QC block involves asking the network (see below) for the creation of one or more entangled pairs. Entanglement generation is a slow and non-deterministic process, with the expected time to generate an entangled pair varying based on the distance between nodes and the hardware used. Generating entanglement on current NV-center hardware takes approximately 1s on average (lab setting) \cite{delle2025operating}. Using anticipated values for trapped ion hardware, the expected entanglement generation time is as low as 15ms on average (lab setting)~\cite{private}. For more details see \Cref{sec:appendix:params}.
\end{itemize}
The programs constituting an application can be developed separately, e.g. a client programming a client program and a server programming a server program.
Each node has its own Qoala runtime, which executes Qoala programs.
Qoala requires nodes to have a \ac{CPS} (executing CL and CC blocks) and a \ac{QPS} (executing QL and QC blocks).

A node can also execute multiple programs concurrently.
This enables better usage of quantum resources, by e.g. executing (part of) one program while another program is waiting for some event \cite{delle2025operating}. 

Qoala assumes the existence of a network control plane (either a central controller or a distributed protocol)~\cite{beauchamp_2025_modularquantumnetworkarchitecture}, which takes an application's demand for entanglement and distributes a network schedule~\cite{skrzypczyk_2021_arch} to the nodes. This schedule consists of sequential time slots specifying when nodes are allowed to use the network to generate entanglement.
Execution of QC blocks (which request entangled pairs) consists of triggering corresponding generation procedures in the allowed time slots. We refer to these slots as \emph{time bins}, and the duration of each slot as the \emph{bin length}.

Qoala extends the QNodeOS architecture~\cite{delle2025operating}, which is also able to (concurrently) execute arbitrary quantum network applications.
QNodeOS treats the classical and quantum parts of the program separately, limiting the implementation of hybrid quantum-classical compilation strategies.
Furthermore, in the current implementation the \ac{CPS} and \ac{QPS} schedulers operate completely independently, preventing advanced scheduling algorithms that oversee execution of the whole program.

Qoala addresses these issues by presenting two new elements.
First, Qoala defines a unified, hybrid quantum-classical executable format for quantum network programs.
The executable format covers both classical and quantum code (in the form of CL, CC, QL, and QC blocks), enabling full-program code analysis and optimization by a compiler.

Second, Qoala specifies how programs are executed.
A single program is \emph{instantiated} one or more times by the user.
For each such program instance, program blocks are converted into \emph{tasks} for the instance. For simplicity, in this work we only consider \emph{blocks}. We assume the scheduler directly executes blocks and do not deal with the underlying runtime mechanism of tasks. 
Blocks are treated as non-preemptable atomic blocks of execution. The scheduler stores these blocks in a graph, which induces precedence and deadline (see below) constraints. When the precedence constraints for a block are satisfied it is considered to be available for execution.

At any point in time, there may be multiple program instances running concurrently in the Qoala runtime of a node.
These instances may be of the same program, different programs, or both.
A \emph{node scheduler} (hereafter just `scheduler') decides at runtime which of the available blocks to execute when, based on its scheduling policy.
When executing multiple instances concurrently, the scheduler may interleave execution of blocks of different instances.

The scheduler oversees execution of both \ac{CPS} and \ac{QPS} blocks.
Blocks may specify \emph{deadlines}, either relative to other blocks or as absolute deadlines.
These deadlines are transferred to their respective blocks at runtime and may be used by the scheduler.
Blocks may also be grouped into \emph{critical sections}, which forces the scheduler to treat such groups as single units of execution.

\subsection{Compilation opportunities}
The features mentioned above (hybrid quantum-classical format with blocks, and a node-wide scheduler) enable particular compilation strategies not possible without the Qoala extension of QNodeOS.
Indeed, a compiler for Qoala programs may apply optimization techniques across classical and quantum blocks.
Moreover, a compiler can guide the scheduler's behavior at runtime:
the compiler may insert deadlines, use critical sections, or use a particular division of code into blocks, in order to increase runtime performance.
These scheduler-guiding compilation strategies are most impactful in scenarios where multiple programs are running concurrently on the node.

While there has not yet been developed a full compilation stack that converts a high-level program (written in e.g. Python) into the Qoala program format,
we already investigate here which different compilation outputs we want to see.

We focus on
\begin{inlinelist}
\item hybrid quantum-classical compilation,
\item different strategies of dividing program code into blocks,
\item different ways of assigning deadlines to blocks,
\item the impact of (not) using critical sections.
\end{inlinelist}

We note that we do this investigation in the context of a quantum internet consisting of autonomous nodes. Therefore, we only consider compilation of single programs (which communicate---using classical communication and entanglement generation---with other programs on other nodes realizing an application), rather than joint compilation of multiple distributed quantum programs.
Furthermore, just like in classical multiprocessing systems, we assume the compiler is not aware of any other programs running on the same node.
It is up to the runtime environment and its scheduler to orchestrate the concurrent execution of programs.
A program compiler can only guide the scheduler as mentioned above.
\subsection{Compilation Strategy Definitions}
\label{sec:background:defs}
\begin{itemize}
    \item \textbf{Hybrid Compilation} refers to the new compilation possibilities made available with the unified Qoala program format. For example, with hybrid compilation it is possible to reorder classical and quantum blocks (as long as data dependencies remain satisfied).
    \item \textbf{Block Compilation}
    \label{sec:background:defs:block-comp}
     determines the number of blocks in a compiled Qoala program. Specific compilation options can be selected to allow for more (or fewer) blocks in a Qoala program.
    \begin{itemize}
        \item \textbf{Selfish Block Compilation} of a Qoala program minimizes the number of quantum local blocks by placing quantum gates in as few blocks as possible. We call this selfish since a program with fewer blocks allows for less interleaving with other programs.
        
        \item \textbf{Cooperative Block Compilation} of a Qoala program increases the total number of quantum local blocks by distributing quantum gates over multiple blocks. We call this cooperative since increasing the number of blocks in a program allows for more opportunities for interleaving. $N$-cooperative block compilation enforces that no more than $N$ quantum gates are in a single block. A block is still allowed to have as many instructions as necessary for loading the quantum memory register and measurement(s). 
    \end{itemize}

    \item \textbf{Deadline Compilation}
    \label{sec:background:defs:deadline}
    After a Qoala program is split into blocks, the compiler can add deadlines. We consider deadline compilation for an \ac{EDF} scheduler. These (soft) deadlines specify the maximum amount of time that should occur between the completion of a block and the next block. 
    \begin{itemize}
        \item \textbf{Deadline Free Compilation}:
        No deadlines are specified. This results in the first available block being executed by the processing units.
         
        \item \textbf{Selfish Deadline Compilation} of a Qoala program assigns the block deadlines such that it is not possible for another block to be inserted between sequential blocks (if multiple are available). This is achieved by setting the deadline equal to the processing time of a single instruction. For quantum blocks the deadline is set to be the \emph{quantum instruction execution time} (see \ref{sec:appendix:params}) and for classical blocks the deadline is set to be the \emph{classical instruction execution time}. We call this selfish since it prevents blocks from other programs from being interleaved.
        
        \item \textbf{Cooperative Deadline Compilation} of a Qoala program increases the deadlines to allow for the interleaving of blocks. We call this cooperative since it allows for more interleaving of blocks between program instances. An $M$-cooperative deadline compilation specifies the deadline of a block to be $M \cdot t_i$, where $t_i$ is the time to complete an instruction of that block type. For quantum blocks we have deadlines set to $M \cdot$(\emph{quantum instruction execution time} + max \emph{gate duration}), and for classical blocks we have deadlines set to $M \cdot$ \emph{classical instruction execution time}.
    \end{itemize}

    \item \textbf{Critical Section Compilation} 
    \label{sec:background:defs:crit} uses critical sections to enforce a group of blocks to be executed as a single atomic unit. In this compilation method, a critical section begins at the final entanglement generation, and ends at the last quantum measurement. All of the blocks (quantum and classical) in between are part of the critical section. This prevents any other program from executing blocks until the final measurement has been made. 
\end{itemize}

\subsection{Evaluation Metrics}
When evaluating application performance we use two different metrics: \emph{execution time} and \emph{success probability}.
Execution time is a classical metric that measures how long it takes an application to run from start to finish. A shorter execution time is better. Success probability is a quantum metric that measures how likely an application is to succeed. An application execution is considered a success when it returns an expected outcome, e.g. a $Z$ basis measurement outcome of 0 when our qubit is expected to be in the $\ket{0}$ state. We run many application executions and take the average in order to determine the overall success probability of the application. A larger success probability is better since it gives us greater confidence in our results.

\begin{figure*}[h]    
    \begin{subfigure}{0.31\linewidth}
        \centering
        \resizebox{1\textwidth}{!}{
            \begin{tikzpicture}[
            l/.style={align=center},
            q/.style={rectangle, draw=purple!50, fill=purple!5, very thick, minimum width=28mm, minimum height = 7mm, align=center},
            miniq/.style={rectangle, draw=purple!50, fill=purple!5, very thick, minimum width=3mm, minimum height = 3mm, align=center},
            minic/.style={rectangle, draw =green!60, fill=green!5, very thick, minimum width = 3mm, minimum height = 3mm, align=center},
            c/.style={rectangle, draw =green!60, fill=green!5, very thick, minimum width = 28mm, minimum height = 7mm, align=center},
            legend/.style={rectangle, draw=black!100, very thick, minimum width = 30mm, minimum height = 10mm, align=left, text width=30mm}
            ]

            \node[q] at (0,0) (B1)  {Initialize $\ket{\psi}$};
            \node[c, below=1mm of B1] (B2) {Receive $\theta_1$};
        
            \node[q, below=1mm of B2] (B3) {$R_X(\theta_1)\ket{\psi}$};
            \node[c, below=1mm of B3] (B4) {Receive $\theta_2$};
            \node[q, below=1mm of B4] (B5) {$R_X(\theta_2)\ket{\psi}$\\Measure $\ket{\psi}$};

            \node[c, right=3mm of B1] (B1') {Receive $\theta_1$};
            \node[c, below=1mm of B1'] (B2') {Receive $\theta_2$};
            \node[q, below=1mm of B2'] (B3')  {Initialize $\ket{\psi}$\\$R_X(\theta_1 + \theta_2)\ket{\psi}$\\Measure $\ket{\psi}$};
        
            \node[l, above=1mm of B1] (LU) {\textbf{Unoptimized}\\ \textbf{Server Program}};
            \node[l, above=1mm of B1'] (LO) {\textbf{Optimized}\\ \textbf{Server Program}};
        
        
        \end{tikzpicture}
        }
        \captionsetup{width=0.95\linewidth}
        \caption{An illustration of the optimizations performed on the rotation application for $n=2$. In the optimized program the classical and quantum blocks have been reordered so that the quantum state $\ket{\psi}$ is not initialized until all of the rotation angles have been received. This minimizes the time the state spends deteriorating in memory. Additionally, the quantum blocks have been combined into a single block to reduce classical processing overhead. The $X$ rotation gates have been combined into a single $X$ rotation, reducing the effects of gate fidelity on the qubit.}
        \label{fig:hybrid-comp-ex}
        \caption*{Fig. 1}
    \end{subfigure}
    \begin{subfigure}{0.68\linewidth}
        \centering
        \resizebox{1\linewidth}{!}{
            \begin{tikzpicture}[
            l/.style={align=center},
            ll/.style={align=right, text width=50mm},
            qc/.style={rectangle, draw=violet!100, fill=violet!15, very thick, minimum width=34mm, minimum height = 7mm, align=center},
            ql/.style={rectangle, draw=purple!50, fill=purple!5, very thick, minimum width=34mm, minimum height = 7mm, align=center},
            miniql/.style={rectangle, draw=purple!50, fill=purple!5, very thick, minimum width=3mm, minimum height = 3mm, align=center},
            miniqc/.style={rectangle, draw=violet!100, fill=violet!15, very thick, minimum width=3mm, minimum height = 3mm, align=center},
            minic/.style={rectangle, draw =green!60, fill=green!5, very thick, minimum width = 3mm, minimum height = 3mm, align=center},
            c/.style={rectangle, draw =green!60, fill=green!5, very thick, minimum width = 34mm, minimum height = 7mm, align=center},
            legend/.style={rectangle, draw=black!100, very thick, minimum width = 40mm, minimum height = 22mm, align=right}
            ]
            \tikzstyle{every node}=[font=\large] 
            \node[qc] at (-0.5,0) (B1)  {Generate \\ Entanglement};
            \node[ql, below=1mm of B1] (B2) {Remote State \\ Preparation $\ket{+_{\theta_i}}$};
            \node[c, below=1mm of B2] (B3)  {Send $Z_\text{correction}$};
            \node[c, below=4mm of B3] (B4)  {Send $\delta_i$};
            \node[c, below=1mm of B4] (B5) {Receive $m_i$};

            \draw[decorate, decoration={brace, amplitude=4mm, mirror}, very thick] (B1.north west) -- (B3.south west) node[midway, left=3.5mm, align=center] {$n$};
             \draw[decorate, decoration={brace, amplitude=4mm, mirror}, very thick] (B4.north west) -- (B5.south west) node[midway, left=3.5mm, align=center] {$n$};       
             
            \node[qc] at (4.5,0) (B1')  {Generate \\ Entanglement};
            \node[c, below=1mm of B1'] (B2') {Receive $Z_{\text{correction}}$};
            \node[ql, below=1mm of B2'] (B3') {$R_Z(Z_{\text{correction}})$};
            
            \node[ql, below=3.5mm of B3'] (B4') {$CZ \ket{+_{\theta_{i}}+_{\theta_{i+1}}}$};
            
            \node[c, below=3.5mm of B4'] (B5') {Receive $\delta_i$};
            \node[ql, below=1mm of B5'] (B6') {$R_Z(\delta_i)$ \\ $m_i$=Measure $\ket{+_{\theta_i}}$};
            \node[c, below=1mm of B6'] (B7') {Send $m_i$};

            \draw[decorate, decoration={brace, amplitude=4mm, mirror}, very thick] (B1'.north west) -- (B3'.south west) node[midway, left=3.5mm, align=center] {$n$};
            
            \draw[decorate, decoration={brace, amplitude=4mm, mirror}, very thick] (B4'.north west) -- (B4'.south west) node[midway, left=3.5mm, align=center] {$n-1$};
            
            \draw[decorate, decoration={brace, amplitude=4mm, mirror}, very thick] (B5'.north west) -- (B7'.south west) node[midway, left=3.5mm, align=center] {$n$};

            \node[qc] at (8.5,0) (B1'')  {Generate \\ Entanglement};
            \node[c, below=1mm of B1''] (B2'') {Receive $Z_{\text{correction}}$};

            \node[ql, below=3.5mm of B2''] (B3'') {$CZ \ket{+_{\theta_{i}}+_{\theta_{i+1}}}$};
            
            \node[c, below=3.5mm of B3''] (B4'') {Receive $\delta_i$};
            \node[ql, below=1mm of B4''] (B5'') {$R_Z(\delta_i + Z_{\text{correction}})$ \\ $m_i$=Measure $\ket{+_{\theta_i}}$};
            \node[c, below=1mm of B5''] (B6'') {Send $m_i$};

            \draw[decorate, decoration={brace, amplitude=4mm}, very thick] (B1''.north east) -- (B2''.south east) node[midway, right=3.5mm, align=center] {$n$};
            \draw[decorate, decoration={brace, amplitude=4mm}, very thick] (B3''.north east) -- (B3''.south east) node[midway, right=3.5mm, align=center] {$n-1$};
            \draw[decorate, decoration={brace, amplitude=4mm}, very thick] (B4''.north east) -- (B6''.south east) node[midway, right=3.5mm, align=center] {$n$};

            \node[l, above=1mm of B1] (C) {\textbf{Client BQC}\\ \textbf{Program}};
            \node[l, above=1mm of B1'] (SU) {\textbf{Unoptimized Server}\\ \textbf{BQC Program}};
            \node[l, above=1mm of B1''] (SO) {\textbf{Optimized Server}\\ \textbf{BQC Program}};
        
            \node[legend, below left=1mm and 52mm of B5''] (legend) {};
            \node[ll, below right=-22mm and -59mm of legend] (clabel) {\textbf{Classical:}};
            \node[minic, right=1mm of clabel] {};
            \node[ll, below=0mm of clabel] (qllabel) {\textbf{Quantum Local:}}; 
            \node[miniql, right=1mm of qllabel] {};
            \node[ll, below=0mm of qllabel] (qclabel) {\textbf{Quantum \\ Communication:}}; 
            \node[miniqc, below right=-4.5mm and 1mm of qclabel] {};
             
        \end{tikzpicture}       
        }
        \captionsetup{width=0.95\linewidth}
        \caption{An illustration of the client and server programs for an $n$ qubit BQC application. Each bracketed group of blocks is repeated the labeled number of times. The client program: (1) generates entanglement with the server, (2) remotely prepares a state on the server (by performing a local measurement on the entangled qubit), (3) sends the correction for the prepared state, (4) sends the desired measurement angle, $\delta_i$, and (5) receives the measurement outcome from the server. The unoptimized server program: (1) generates entanglement with the client, (2) receives the correction, (3) corrects the prepared state, (4) applies $CZ$ gates to the prepared states (if $n>1$), (5) receives the measurement angle, $\delta_i$, (6) measures the prepared state, and (7) sends the measurement outcome to the client. In the optimized server program the classical and quantum local operations are reordered, which allows the $R_Z$ gates to be combined. }
        \label{fig:hybrid-bqc-ex}
    \end{subfigure}
    \label{fig:hybrid-comp}
\end{figure*}

\section{Simulation Configuration}
\label{sec:sim-details}

For simulating the execution of applications on quantum processing nodes in a quantum network we make use of the simulator used in the Qoala paper \cite{vandervecht2025qoala,qoalasim}. This simulator allows for low level details when simulating the timings between different components in the architecture, such as the classical/quantum instruction execution time, shared memory processing time, classical communication latency, quantum gate duration, and more. For a detailed description see \ref{sec:appendix}.

Our evaluation parameters are based on real and anticipated values for trapped-ion hardware \cite{private}. The default parameters for our trapped-ion hardware configuration are shown in Table~\ref{tbl-param}. 

We make the following simplifying assumptions: \begin{inlinelist}
\item gates operating on the same number of qubits have identical gate duration and fidelity,
\item a gate set of \{$X$, $Z$, $H$, $R_X(\theta)$, $R_Z(\theta)$, $R_Y(\theta)$, $CZ$\} is supported,
\item gates can be applied between any pair of qubits, and
\item all qubits can be used for entanglement generation.
\end{inlinelist}

We model entanglement generation as a stochastic process where each attempt takes a fixed amount of time, and succeeds with some probability. We average our simulation results over 10 different initial random seeds generated using a publicly available quantum random number generator \cite{Camacho_2023_LabQRNG}. 

\begin{table}[]
\begin{center}
\centering
\caption{Default trapped-ion hardware values.
}
\label{tbl-param}
\begin{tabular}{|c|c|}
\hline
\textbf{Parameter }                                                                     & \textbf{Value}      \\ \hline
Single qubit gate duration                                                     & 26.6$\mu$s \cite{avis_2023_reqprocnoderepeater} \\ \hline
Two qubit gate duration                                                        & 107$\mu$s \cite{krutyanskiy_2023_trappedionrepeater} \\ \hline
Single qubit gate fidelity                                                     & 0.99  \cite{avis_2023_reqprocnoderepeater}     \\ \hline
Two qubit gate fidelity                                                        & 0.95  \cite{krutyanskiy_2023_trappedionrepeater}     \\ \hline
Dephasing ($T_2$) time                                                            & 10s    \cite{private}    \\ \hline
\begin{tabular}[c]{@{}l@{}}Quantum instruction \\ execution time\end{tabular}  & 50$\mu$s \cite{delle2025operating} \\ \hline
\begin{tabular}[c]{@{}l@{}} Expected entanglement \\ generation time (lab setting) \end{tabular}  & 15.26ms \cite{private}    \\ \hline
\begin{tabular}[c]{@{}l@{}} Entanglement generation \\ probability (lab setting)\end{tabular}  & 0.013  \cite{private}  \\ \hline
\begin{tabular}[c]{@{}l@{}} Duration of entanglement \\ generation attempt (lab setting)\end{tabular}  & 200 $\mu s$  \cite{private}  \\ \hline
\begin{tabular}[c]{@{}l@{}}Entangled pair fidelity\end{tabular}  & 0.95  \cite{private}  \\ \hline
\begin{tabular}[c]{@{}l@{}} Classical communication \\ latency (lab setting) \end{tabular}      & 0.155ms \cite{bovy_2002_internetdelay} \\ \hline

\end{tabular}
\end{center}
\end{table}

\section{Hybrid Quantum-Classical Compilation}
\label{sec:hybrid-comp}

As mentioned in \ref{sec:background:qoala-recap}, the Qoala program format enables hybrid quantum-classical compilation. This allows for the reordering and combining of blocks. 

\subsection{Rotation Application}
\emph{Can hybrid compilation improve program performance compared to QNodeOS without hybrid compilation?}
To explore this, we first propose a simple client-server application where hybrid compilation provides a clear advantage. We call this example the \emph{rotation application}. 
Our client program sends $n$ angles over a classical communication channel to the server. These angles are chosen at random with the constraint that the angles sum to $2\pi$. Our server program performs $R_X$ gates on a single qubit based on the measurement angles received.

The unoptimized server program, referred to as $S_{unopt}$, will: 
\begin{itemize}
    \item [1.] Initialize the qubit to one of the six $\pm \{X,Y,Z\}$ basis states (chosen uniformly at random).
    \item [2.] Receive an angle $\theta_i$ from the client and will perform an X rotation by $\theta_i$, repeating this processing $n$ times
    \item [3.] Measure the qubit in the same basis it was initialized.
\end{itemize}

Using hybrid compilation, the classical and quantum blocks in the program can be reorganized. This allows for the quantum blocks to be combined. 

The resulting optimized server program, called $S_{opt}$, will:
\begin{itemize}
    \item [1.] Receive an angle $\theta_i$ from the client. Repeat this $n$ times.
    \item [2.] Sum the received angles $\theta$ = $\sum_i \theta_i$.
    \item [3.] Initialize the qubit to one of the six $\pm \{X,Y,Z\}$ basis states (chosen uniformly at random).
    \item [4.] Perform an X rotation by $\theta$.
    \item [5.] Measure the qubit in the same basis it was initialized.
\end{itemize}
An illustration of the difference between $S_{unopt}$ and $S_{opt}$ can be seen in Fig. \ref{fig:hybrid-comp-ex}.

Notably, in $S_{opt}$, there is a single rotation gate, whereas in $S_{unopt}$ the number of rotation gates scales with $n$. When considering imperfect quantum gates (fidelity $<$ 1), we expect $S_{opt}$ to have a higher success probability than $S_{unopt}$ due to executing fewer (noisy) gates.

\subsubsection{Simulation Configuration}
\label{sec:hybrid-comp:rot-eval}
We separately vary the single qubit gate fidelity and the classical communication latency, comparing how they effect the unoptimized and optimized rotation application variants. For single qubit gate fidelity we sweep over values of $\{0.95, 0.96, 0.97, 0.98, 0.99, 0.995, 0.999\}$. In order to isolate the effect of gate fidelity on success probability, we set all other parameters to the ideal scenario (infinite $T_2$ time). When varying the classical communication latency we choose values that are a fraction of the $T_2$ time $\{0.1,0.2,0.3,0.4,0.5,0.6,0.7,0.8,0.9,1.0\}$ in order to see the difference in memory decoherence between $S_{unopt}$ and $S_{opt}$. We run each program 1000 times and plot the average over our 10 initial seeds. 

\subsubsection{Results}

\begin{figure}
    \centering
    \includegraphics[width=\linewidth]{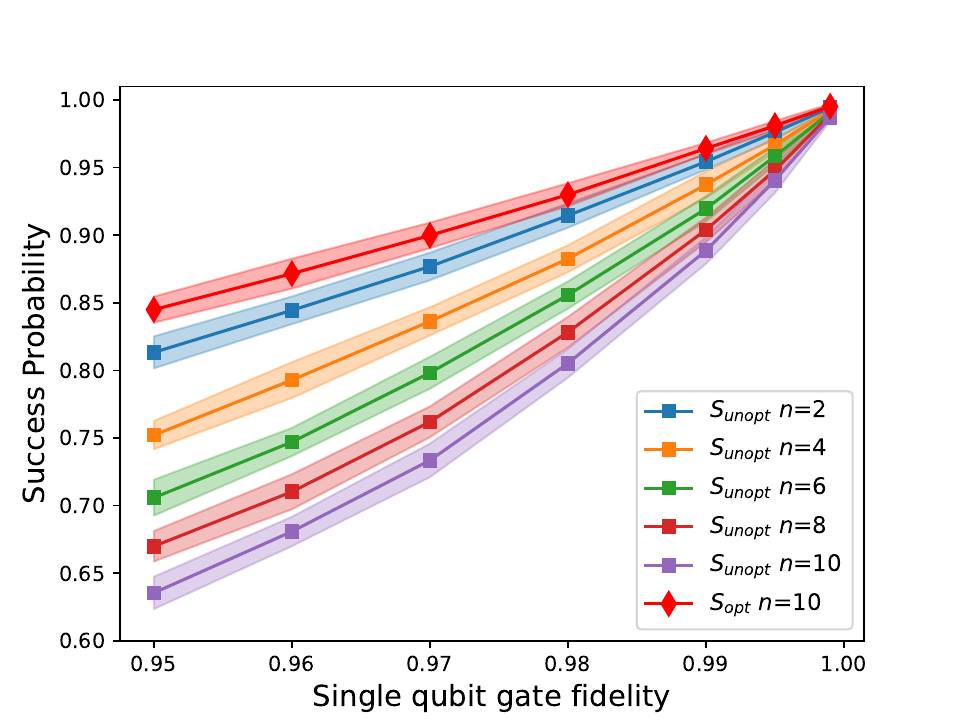}
    \caption{Success probability of the rotation application for varied single qubit gate fidelities. $S_{opt}$'s success probability is independent of $n$, the number of angles sent, hence a single line is shown. 
    }
    \label{fig:rotation-gatefid-succprob}
\end{figure}

 In Fig. \ref{fig:rotation-gatefid-succprob} we see that $S_{opt}$ outperforms $S_{unopt}$ when gate fidelities are imperfect. The success of $S_{opt}$ is independent of $n$, the number of angles sent. This independence is due to $S_{opt}$ only performing a single R$_X$ gate, regardless of $n$. As $n$ increases, so does the improvement in the success probability of $S_{opt}$, with increases of 1.86\%, 5.70\%, 9.35\%, 12.85\%, and 16.11\% for $n=2, 4, 6, 8, 10$, respectively.

\begin{figure}
    \centering
    \includegraphics[width=\linewidth]{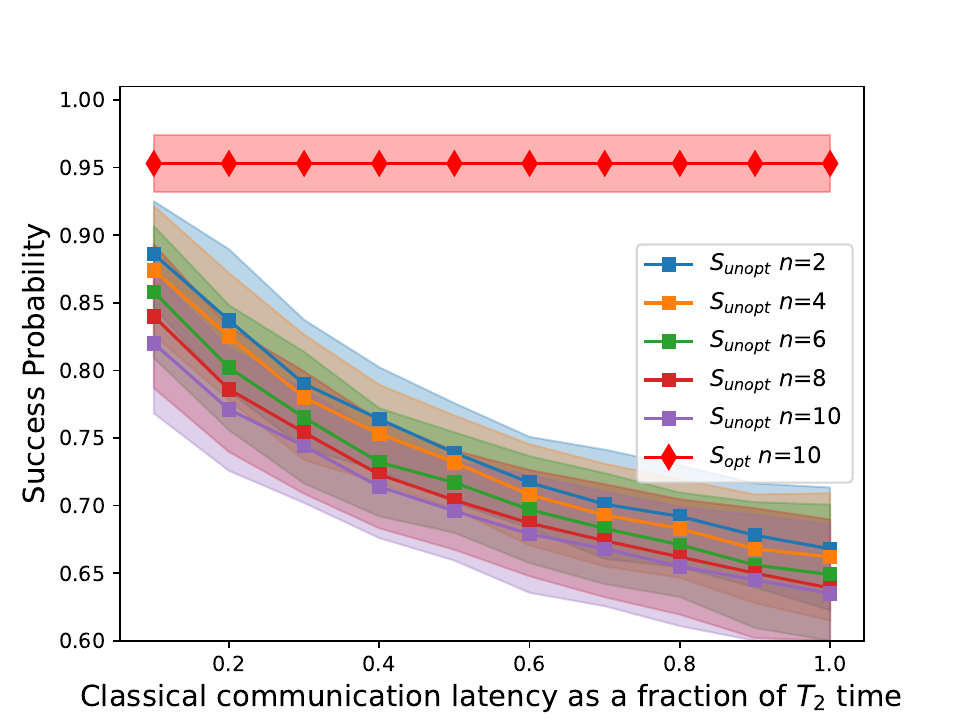}
    \caption{Success probability of the rotation for varied classical communication latencies. $S_{opt}$'s success probability is independent of $n$, the number of angles sent, hence a single line is shown. 
    }
    \label{fig:rotation-cc-succprob}
\end{figure}

In Fig. \ref{fig:rotation-cc-succprob} we see that $S_{opt}$'s success probability is unaffected by the classical communication latency. This is due to $S_{opt}$ receiving all the measurement angles before initializing the qubit. In $S_{unopt}$, the qubit is stored in memory while waiting for classical communications to be received. As the classical communication latency approaches the $T_2$ time we see the success probability decrease. Varying the classical communication latency has an equal effect on the execution time of $S_{opt}$ and $S_{unopt}$ since they receive an equal number of classical communications.

\begin{table}[]
        \centering
        \caption{Distances and hop counts between nodes in the SURF network \cite{rabbie_2022_surfnetinfra, rabbie_2020_surfnetrepo}.}
        
        \label{tbl:distances} 
            \begin{tabular}{|c|c|c|c|}
        \hline
        \textbf{Server}  & \textbf{Client}     & \textbf{Distance (km)} & \textbf{Hops} \\ \hline
        Delft 1 & Delft 1        & 0             & 0    \\ \hline
        Delft 1 & Delft 2    & 2.2           & 0    \\ \hline
        Delft 1 & Rotterdam 1    & 16.8          & 1    \\ \hline
        Delft 1 & Den Haag 2 & 19.8          & 0    \\ \hline
        Delft 1 & Den Haag 1 & 26.3          & 1    \\ \hline
        Delft 1 & Leiden 1   & 30.6          & 0    \\ \hline
        Delft 1 & Rotterdam 2    & 33.1          & 0    \\ \hline
        Delft 1 & Rotterdam 3    & 40.2          & 1    \\ \hline
        Delft 1 & Leiden 2   & 47.9          & 2    \\ \hline
        Delft 1 & Leiden 3   & 55.2          & 3    \\ \hline
        \end{tabular}
\end{table}

\subsection{BQC Application}

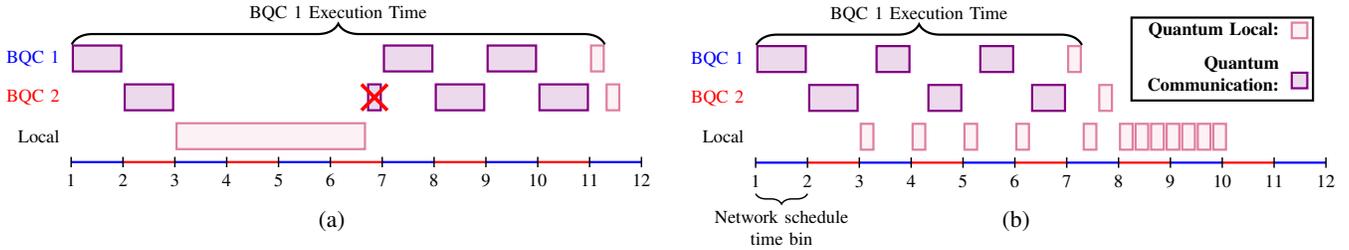
\begin{figure*}[]
    \centering
    \begin{subfigure}{0.495\textwidth}
        \resizebox{1\linewidth}{!}{
            \begin{tikzpicture}[node distance = 1cm,
            background rectangle/.style={fill=white},
            ql/.style={rectangle,
                            draw=purple!50, 
                            fill=purple!5, 
                            very thick, 
                            align=center, 
                            minimum height = 0.5cm,
                            },
            qc/.style={rectangle,
                            draw=violet!100, 
                            fill=violet!15, 
                            very thick, 
                            align=center, 
                            text width=0.7cm, 
                            minimum height = 0.5cm,
                            },
            cross/.style={cross out, 
                            draw, 
                            minimum size=2*(#1-\pgflinewidth), 
                            inner sep=0pt,
                            outer sep=0pt}]

    \foreach \i in {0,...,5} {
        \draw[very thick, color=blue] ($(\i,0) + (\i,0)$) -- ($(\i,0) + (\i,0) + (1,0)$);
    }
    \foreach \i in {0,...,4} {
        \draw[very thick, color=red] ($(\i,0) + (\i,0) + (1,0)$) -- ($(\i,0) + (\i,0) + (2,0)$);
    }
    
    \foreach \i in {1,...,12} {
        \draw[thick] ($(\i,0.1) - (1,0)$) -- ($(\i,-0.1) - (1,0)$) node [below] {\i};
    }
 
    \coordinate (L) at (0,0.5);
    \coordinate (BQC1) at (0,2.0);
    \coordinate (BQC2) at (0,1.25);
   \node[qc] (P1_1) at ($(0.5,0)+(BQC1)$) {};
   \node[qc] (P1_2) at ($(6.5,0)+(BQC1)$) {};
   \node[qc] (P1_3) at ($(8.5,0)+(BQC1)$) {};
   \node[ql, text width = 0.1mm] (P1_4) at ($(10.15,0)+(BQC1)$) {};

   \node[qc] (P2_1) at ($(1.5,0)+(BQC2)$) {};
   \node[qc, text width = 0.1mm] (P2_2) at ($(5.85,0)+(BQC2)$) {};
   \node[qc] (P2_3) at ($(7.5,0)+(BQC2)$){};
   \node[qc] (P2_4) at ($(9.5,0)+(BQC2)$) {};
   \node[ql, text width = 0.1mm] (P2_5) at ($(10.45,0)+(BQC2)$) {};

    \node[cross, color=red, line width=0.7mm, minimum width = 5mm, minimum height = 5mm] at (P2_2) {};

   \node[ql, text width = 3.4cm] (L1_1) at ($(3.85,0)+(L)$) {};

    \draw[decorate, decoration={brace, amplitude=4mm},very thick] (P1_1.north west) -- (P1_4.north east) node[above=3mm, midway] {BQC 1 Execution Time};

    \node (label3) at ($(-0.1,0 |- L1_1.center)$) [left] {Local};
    \node (label2) at ($(-0.1,0 |- P1_1.center)$) [left, text=blue] {BQC 1};
    \node (label1) at ($(-0.1,0 |- P2_1.center)$) [left,text=red] {BQC 2};

\end{tikzpicture}
        }
        \caption{}
        \label{fig:block-comp-diagram-a}
    \end{subfigure}
    \begin{subfigure}{0.495\textwidth}
        \resizebox{1\linewidth}{!}{
            \begin{tikzpicture}[node distance = 1cm,
            background rectangle/.style={fill=white},
            ql/.style={rectangle,
                            draw=purple!50, 
                            fill=purple!5, 
                            very thick, 
                            align=center, 
                            minimum height = 0.5cm,
                            },
            qc/.style={rectangle,
                            draw=violet!100, 
                            fill=violet!15, 
                            very thick, 
                            align=center, 
                            text width=0.7cm, 
                            minimum height = 0.5cm,
                            },l/.style={align=center},
            ll/.style={align=right, 
                            text width=47mm},
            miniql/.style={rectangle, 
                            draw=purple!50, 
                            fill=purple!5, 
                            very thick, 
                            minimum width=3mm, 
                            minimum height = 3mm, 
                            align=center},
            miniqc/.style={rectangle, 
                            draw=violet!100, 
                            fill=violet!15, 
                            very thick, 
                            minimum width=3mm, 
                            minimum height = 3mm, 
                            align=center},
            minic/.style={rectangle, 
                            draw =green!60, 
                            fill=green!5, 
                            very thick, 
                            minimum width = 3mm, 
                            minimum height = 3mm, 
                            align=center},
            c/.style={rectangle, 
                            draw =green!60, fill=green!5, very thick, minimum width = 31mm, minimum height = 7mm, align=center},
            legend/.style={rectangle, 
                            draw=black!100, 
                            very thick, 
                            minimum width = 35mm, 
                            minimum height = 16mm, 
                            align=right},]

    \foreach \i in {0,...,5} {
        \draw[very thick, color=blue] ($(\i,0) + (\i,0)$) -- ($(\i,0) + (\i,0) + (1,0)$);
    }
    \foreach \i in {0,...,4} {
        \draw[very thick, color=red] ($(\i,0) + (\i,0) + (1,0)$) -- ($(\i,0) + (\i,0) + (2,0)$);
    }
    
    \foreach \i in {1,...,12} {
        \draw[thick] ($(\i,0.1) - (1,0)$) -- ($(\i,-0.1) - (1,0)$) node [below] {\i};
    }
    
    \coordinate (L) at (0,0.5);
    \coordinate (BQC1) at (0,2.0);
    \coordinate (BQC2) at (0,1.25);
   \node[qc,text width = 0.7cm] (P1_1) at  ($(0.5,0) +(BQC1)$){};
   \node[qc, text width = 0.4cm] (P1_2) at ($(2.65,0)+(BQC1)$)  {};
   \node[qc, text width = 0.4cm] (P1_3) at ($(4.65,0)+(BQC1)$)  {};
   \node[ql, text width = 0.1mm] (P1_4) at ($(6.15,0)+(BQC1)$)  {};

   \node[qc, text width = 0.7cm] (P2_1) at ($(1.5,0)+(BQC2)$) {};
   \node[qc, text width = 0.4cm] (P2_2) at ($(3.65,0)+(BQC2)$) {};
   \node[qc, text width = 0.4cm] (P2_3) at ($(5.65,0)+(BQC2)$) {};
   \node[ql, text width = 0.1mm] (P2_4) at ($(6.75,0)+(BQC2)$) {};

   \node[ql, text width = 0.1mm] (L1_1) at  ($(2.15,0)+(L)$) {};
   \node[ql, text width = 0.1mm] (L1_2) at  ($(3.15,0)+(L)$) {};
   \node[ql, text width = 0.1mm] (L1_3) at  ($(4.15,0)+(L)$) {};
   \node[ql, text width = 0.1mm] (L1_4) at  ($(5.15,0)+(L)$) {};
   \node[ql, text width = 0.1mm] (L1_5) at  ($(6.45,0)+(L)$) {};
   \node[ql, text width = 0.1mm] (L1_6) at  ($(7.15,0)+(L)$) {};
   \node[ql, text width = 0.1mm] (L1_7) at  ($(7.45,0)+(L)$) {};
   \node[ql, text width = 0.1mm] (L1_8) at  ($(7.75,0)+(L)$) {};
   \node[ql, text width = 0.1mm] (L1_9) at ($(8.05,0)+(L)$) {};
   \node[ql, text width = 0.1mm] (L1_10) at ($(8.35,0)+(L)$) {};
   \node[ql, text width = 0.1mm] (L1_11) at ($(8.65,0)+(L)$) {};
   \node[ql, text width = 0.1mm] (L1_12) at  ($(8.95,0)+(L)$) {};
   

    \draw[decorate, decoration={brace, amplitude=4mm},very thick] (P1_1.north west) -- (P1_4.north east) node[above=3mm, midway] {BQC 1 Execution Time};
    
    \draw[decorate, decoration={brace, mirror, amplitude=2mm}, thick] (0,-0.6) -- (1,-0.6) node [midway, below=2mm, align=center] {Network schedule \\ time bin};

    \node (label3) at ($(-0.1,0 |- L1_1.center)$) [left] {Local};
    \node (label2) at ($(-0.1,0 |- P1_1.center)$) [left, color=blue] {BQC 1};
    \node (label1) at ($(-0.1,0 |- P2_1.center)$) [left,color=red] {BQC 2};

   \node[legend] at (9,2.0) (legend) {};
    \node[ll, below right=-16.5mm and -55mm of legend] (qllabel) {\textbf{Quantum Local:}}; 
    \node[miniql, right=1mm of qllabel] {};
    \node[ll, below=1mm of qllabel] (qclabel) {\textbf{Quantum \\ Communication:}}; 
    \node[miniqc, below right=-5mm and 1mm of qclabel] {}; 
\end{tikzpicture}
        }
        \vspace{-1.2cm}
        \caption{}
        \label{fig:block-comp-diagram-b}
    \end{subfigure}
    
    \caption{Shown here is a simplified execution of quantum blocks on the server QPS for two 3 qubit BQC applications and a local program. In (a) the local program is compiled block selfishly, and in (b) the local program is compiled block cooperatively. The network schedule is depicted on the x axis, and the length of each time bin is the expected time to generate a single entangled pair. The color corresponds to which BQC application is allowed to generate entanglement during that bin. 
    In (a) the local program prevents both BQC applications from generating entanglement and increases their execution times. In the time bin from 6-7 the second BQC application fails to generate entanglement due to not enough time remaining in the bin. 
    In (b), block cooperative compilation of the local program allows it to be interleaved with the entanglement generation of the BQC applications. The execution times for BQC are reduced at the cost of increasing the execution time of the local program.  }
    \label{fig:block-comp-diagram}
\end{figure*}

\emph{Can hybrid compilation provide an advantage for more realistic applications?} We chose to evaluate BQC, since it is commonly seen as a future application for quantum networks and has been demonstrated in experimental settings \cite{barz_2012_bqcdemo, Drmota_2024_trapionBQC, lukin_2024_bqcdemo}, including from QNodeOS \cite{delle2025operating}.

See Fig. \ref{fig:hybrid-bqc-ex} for an overview of the BQC protocol and program structure. In our unoptimized BQC server program, two $Z$ rotations are executed on each qubit. One to correct the qubit after remote state preparation, and one before measuring the qubit. In our optimized version of the server program, these two $Z$ rotations are combined into a single rotation by adding the rotation angles, similar to the optimization for $S_{opt}$.

\subsubsection{Simulation Configuration} 
We evaluate BQC applications with a varying number of qubits, $n=\{3,5\}$. We examine the effects of varying the single qubit gate fidelity, and the distance between the server and client. When varying the single qubit gate fidelity we sweep over values of $\{0.95, 0.96, 0.97, 0.98, 0.99, 0.995, 0.999\}$, and set all other parameters to be perfect (entanglement fidelity of 1.0, two qubit gate fidelity of 1.0, infinite $T_2$ time). 
Varying the distance between the client and the server modifies two parameters: the classical communication latency, and the time to generate entanglement. We use a real network topology, SURFnet \cite{rabbie_2022_surfnetinfra, rabbie_2020_surfnetrepo}, a research and education network in The Netherlands, and consider placing our server and client at different locations in the network as shown in Table \ref{tbl:distances}. We calculate the entanglement generation time as a function of the distance, and the classical communication time as a function of the distance and hop count. For more details, see \cref{sec:appendix:params}.

\subsubsection{Results}
For BQC there is very little increase in success probability for the optimized program when varying the distance (0.42\% $n=3$, 0.86\% $n=5$ on average). The dominating sources of noise are the fidelities of two qubit gates (0.95), and entangled pairs (0.95). When isolating single qubit gate fidelity, and considering all other noise sources to be ideal, there is a larger improvement for the optimized program (0.63\% $n=3$, 1.43\% $n=5$ on average). 
Similarly, time to generate entanglement is the dominating factor for execution time. The reduction in number of single qubit gates (76.6$\mu$s per gate) in the optimized program results in a negligible decrease in execution time when varying the distance (0.52\% $n=3$, 0.63\% $n=5$ on average).

\section{Block Compilation}
\label{sec:block-comp}

As defined in \ref{sec:background:defs:block-comp}, block compilation allows programs to influence the scheduler by increasing or reducing the number of blocks their code is split between. We explore three different scenarios, identifying benefits and trade-offs of this compilation method.

\subsection{Scenario 1}
\label{sec:block-comp:scen1}
We consider a single server $S$, with $c$ clients $C_1, \dots C_c$, where each $(S,C_i)$ pair runs its own BQC application. Each client runs a single BQC client program, and the server runs $c$ BQC server programs. Additionally, the server runs a single local program which does the following:
\begin{itemize}
    \item[1.] Initialize a qubit to $\ket{0}$.
    \item[2.] Execute 8 single qubit gates on the qubit.
    \item[3.] Measure the qubit.
    \item[4.] Repeat steps 1-3 200 times.
\end{itemize}

The execution time of this local program is approximately 3.5 times the expected duration of generating entanglement (for anticipated trapped ion hardware parameters \cite{private}). In the selfishly block compiled version of the local program, all four steps are contained in a single quantum block, preventing the BQC applications from interleaving (see Fig. \ref{fig:block-comp-diagram-a}). We expect the execution of the selfishly block compiled local program to increase the execution time of the BQC applications by delaying their entanglement generation attempts.

Using 8-cooperative block compilation on the local program, each iteration of steps 1-3 will be contained in a separate block. This enables the BQC applications to interleave entanglement generation while the local program is being executed (see Fig. \ref{fig:block-comp-diagram-b}). We expect this to increase the execution time of the local program, and decrease the execution time of the BQC applications. We do not expect the success probability of the local program to be affected since the qubit is measured and reset in each block.

\begin{table*}[]
    \centering
    \caption{Scenario 1 execution time impacts of compiling the local program 8-block cooperatively vs. block selfishly. 
    }
    \label{tbl:block-comp:scen1}
    \resizebox{0.75\textwidth}{!}{
        \begin{tabular}{|c|c|c|c|c|}
            \hline
            \multirow{2}{*}{\shortstack{\\[1pt] Number of \\ clients ($c$)}} & \multicolumn{2}{c|}{\rule{0pt}{2.5mm} BQC Program Size $n=3$} & \multicolumn{2}{c|}{\rule{0pt}{2.5mm} BQC Program Size $n=5$} \\
            \cline{2-5} 
             &  \begin{tabular}[c]{@{}c@{}} \rule{0pt}{2.5mm} BQC Execution \\ Time Decrease \end{tabular}  
             &  \begin{tabular}[c]{@{}c@{}} \rule{0pt}{2.5mm} Local Execution \\ Time Increase \end{tabular}  
             &  \begin{tabular}[c]{@{}c@{}} \rule{0pt}{2.5mm} BQC Execution \\ Time Decrease \end{tabular} 
             &  \begin{tabular}[c]{@{}c@{}} \rule{0pt}{2.5mm} Local Execution \\ Time Increase \end{tabular}  \\ 
            \hline
             2 & 29.53\% $\pm$ 5.95\% & 191.30\% $\pm$ 30.21\% & 23.58\% $\pm$ 2.06\% & 313.24\% $\pm$ 24.55\% \\ 
            \hline
             3 & 20.59\% $\pm$ 6.00\% & 230.20\% $\pm$ 59.90\% & 17.21\% $\pm$ 2.03\% & 397.04\% $\pm$ 65.47\%\\
            \hline
            4 & 15.93\% $\pm$ 5.73\% & 256.84\% $\pm$ 86.98\% & 13.60\% $\pm$ 1.98\% & 465.26\% $\pm$ 108.20\%\\
            \hline
            5 & 13.08\% $\pm$ 5.26\% & 280.38\% $\pm$ 110.27\% & 11.51\% $\pm$ 1.89\% & 518.41\% $\pm$ 153.44\% \\
            \hline
        \end{tabular}
    }
\end{table*}

\subsubsection{Simulation Configuration}
\label{sec:block-comp:scen1:sim-config}
We fix our parameters (see Table \ref{tbl-param}), and vary the length of time bins in the network schedule. Bin length is varied as a multiple of the expected entanglement generation time, with lengths of $\{1,2,3,4,5,6,7,8,9,10\}$ (times the expected generation time). Time bins follow a fixed repeating pattern that is randomly permuted for each experiment execution. The number of clients, $c$, is also varied from $\{2,3,4,5\}$. The server node has sufficient quantum memories to execute all of its programs concurrently.
Each simulation run executes the experiment 100 times. We plot the average over 10 simulation runs. 

\subsubsection{Results}
\label{sec:block-comp:scen1:results}

\begin{figure}
    \centering
    \includegraphics[width=\linewidth]{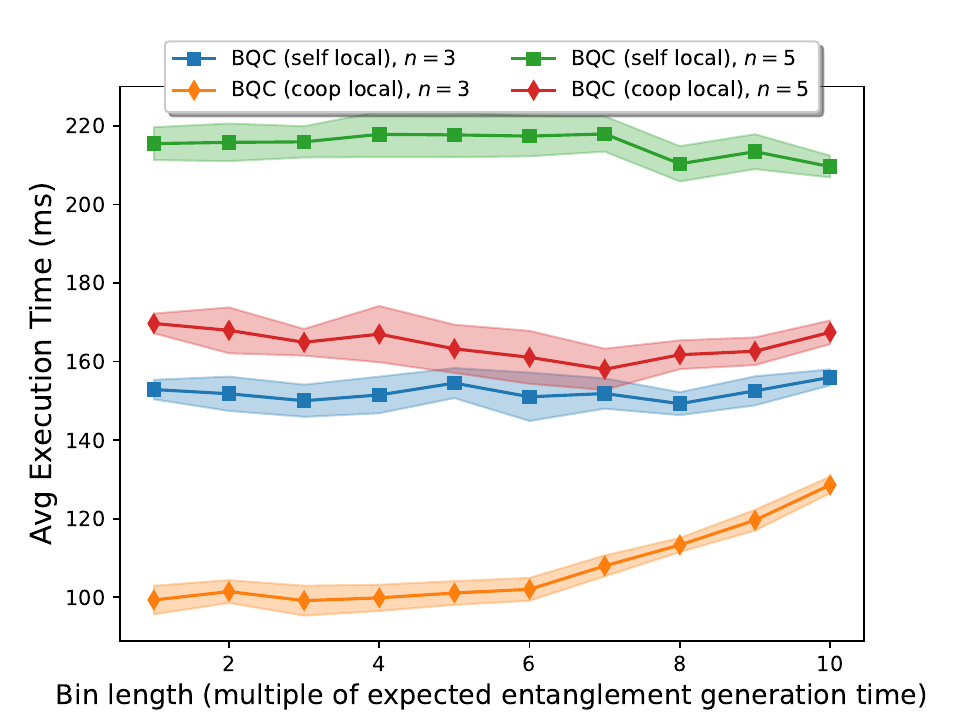}
    \caption{Block compilation Scenario 1 execution time results for two clients each executing a BQC application with a single server, for varied network schedule bin lengths. The server executes a single instance of local program with varying compilation strategies (selfish block, vs. 8-cooperative block). The BQC applications are compiled block selfishly. }
    \label{fig:block-scen1-execution time}
\end{figure}

In Fig. \ref{fig:block-scen1-execution time}, we show the execution time of the BQC application for Scenario 1 with 2 clients. Compiling the local program cooperatively results in a reduced execution time for the BQC applications, 29.53\% for $n=3$, and 23.58\% for $n=5$. This is due to the local program no longer preventing the BQC applications from generating entanglement (as we hypothesized in Fig. \ref{fig:block-comp-diagram}). This improvement for the BQC applications comes at the cost of an increased execution time for the local program. The interleaving of entanglement generation with the local program increases the execution time of the local program by 191.30\% for $n=3$, and 313.24\% for $n=5$. As we increase the number of clients, the reduction in execution time for the BQC applications decreases (see Table \ref{tbl:block-comp:scen1}). Conversely, as the number of clients increases, the increase in execution time of the local program increases.

Our $T_2$ times are significantly larger than the application execution times ($T_2 = 10$s, execution time $<$1s), so success probability (for BQC or the local program) is unaffected when varying the compilation strategy for the local program.

For $n=3$ when running BQC with a cooperatively block compiled local program we see the execution time increase once the bin length exceeds 6. For a bin of length 6, it is highly likely ($\approx$93\%) that each BQC application will complete within a single time bin. The other BQC application must wait until its time bin before attempting entanglement generation. As bin length increases past 6, the BQC applications waiting time increases, resulting in longer execution times for BQC.

\subsection{Scenario 2}
\label{sec:block-comp:scen2}

We now make a slight modification to the local program seen in Scenario 1 (see \ref{sec:block-comp:scen1}). Steps 1 and 2 are repeated 200 times, but only a single measurement is performed at the end of the program. We expect the impact on execution times to be identical to Scenario 1 (see \ref{sec:block-comp:scen1:results}). However, we now expect the success probability of the local program to decrease when compiled 8-block cooperatively due to the qubit staying live in memory for the duration of the BQC applications.

\subsubsection{Simulation Configuration}
\label{sec:block-comp:scen2:sim-config}
Single qubit gate fidelity is set to 1 in order to isolate the effects of decoherence on our local program. The rest is identical to Scenario 1 (see \ref{sec:block-comp:scen1:sim-config}).

\subsubsection{Results}
\label{sec:block-comp:scen2:results}

\newcommand{\pct}[2]{#1\% $\pm$ #2\%}
\begin{table}[]
    \centering
    \caption{Scenario 2 success probability impacts of compiling the local program 8-block cooperatively vs. block selfishly. 
    }
    \label{tbl:block-comp:scen2}
    \begin{tabular}{|c|c|c|}
         \hline
         \multirow{2}{*}{\shortstack{\rule{0pt}{2.5mm} Number of \\ clients ($c$)}} &  \multicolumn{2}{c|}{\shortstack{\rule{0pt}{2.5mm}Local Success \\ Probability Decrease}}
         \\
         
         \cline{2-3}
         & \begin{tabular}[c]{@{}c@{}} \rule{0pt}{2.5mm} BQC Program \\ Size ($n=3$)  \end{tabular}  
         & \begin{tabular}[c]{@{}c@{}} \rule{0pt}{2.5mm} BQC Program \\ Size ($n=5$)  \end{tabular}   \\

        \hline
        2 & \pct{4.31}{1.28} & \pct{7.27}{0.50} \\
        \hline
        3 & \pct{4.93}{1.77} & \pct{8.58}{1.44} \\
        \hline
        4 & \pct{5.26}{2.16} & \pct{10.33}{2.36} \\
        \hline
        5 & \pct{5.83}{2.56} & \pct{10.93}{2.69} \\
        \hline
    \end{tabular}
    
\end{table}

Trends for the execution time are identical to Scenario 1. This is expected since the only difference is where measurements are placed in the local program. The local program experiences a decrease in success probability when compiled block cooperatively (see Table \ref{tbl:block-comp:scen2}), due to the qubit remaining in noisy quantum memory for a longer duration.

\subsection{Scenario 3}
\label{sec:block-comp:scen3}

We now eliminate the local program, and consider a single server, with $c$ clients, each running a single BQC application. We compare compiling all of the server BQC programs using selfish block compilation vs. 1-cooperative block compilation. The individual blocks in the BQC programs are significantly smaller than the expected time for entanglement generation (roughly 1\% or less), so we do not expect to see a decrease in execution time when compiling cooperatively. In fact, we expect a small increase in execution time due to overhead of adding additional blocks. This overhead is due to instructions for loading quantum memory (50$\mu$s per instruction) in each additional quantum block.

\subsubsection{Simulation Configuration}
\label{sec:block-comp:scen3:sim-config}
Same as Scenario 1 (\ref{sec:block-comp:scen1:sim-config}).
\subsubsection{Results}
\label{sec:block-comp:scen3:results}
As expected, compiling the BQC applications on the server block cooperatively vs. block selfishly has a negligible effect on the execution time and success probability. There is a very slight increase in execution time (0.20\% $n=3$, 0.77\% $n=5$, 5 clients), however, this is within the error bars. We observe similar results for 2, 3, and 4 clients.

\section{Deadline Compilation}
\label{sec:deadline-comp}

We consider a single server with $c$ clients, $C_1, \dots C_c$, where each $(S,C_i)$ pair runs a single BQC application. We envision this scenario like an HPC (high performance computing) environment, where the server is the HPC. We compare the performance of user $C_1$'s BQC application with the average performance of users $C_2$, \dots, $C_{c}$. We fix the server compilation strategy of users $C_2$, \dots, $C_{c}$ to be 100-cooperative deadline compilation. If user $C_1$ has the default HPC priority (e.g. student access) their program will be compiled on the server with 100-cooperative deadline compilation. However, if user $C_1$ is a high priority user (e.g. research lab) their program is compiled on the server using selfish deadline compilation. We expect when all users compile with 100-cooperative deadline compilation, there will be no difference in performance between $C_1$ and the other users. When $C_1$ uses selfish deadline compilation, we expect their BQC application to have a shorter execution time than the other users since the execution of their program blocks will be prioritized.

\subsubsection{Simulation Configuration}
\label{sec:deadline-comp:sim-config}
Our configuration is identical to that of Block Compilation Scenario 1 (see \ref{sec:block-comp:scen1:sim-config}).

\subsubsection{Results}
\label{sec:deadline-comp:results}

\begin{figure}
    \centering
    \includegraphics[width=\linewidth]{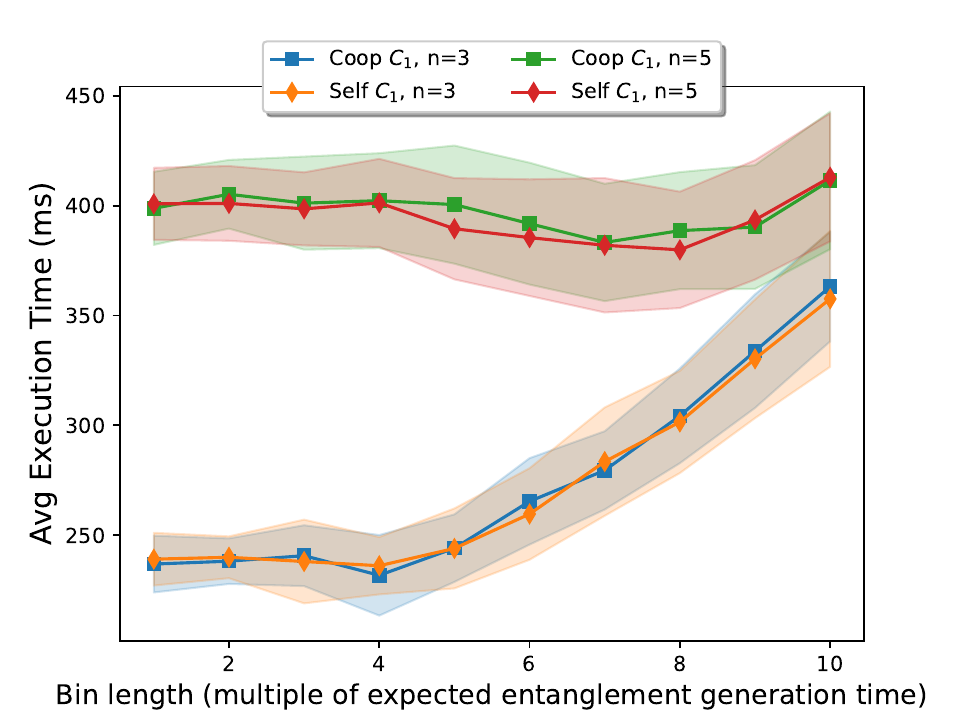}
    \caption{Deadline compilation execution time results for user $C_1$'s BQC application with a total of 5 clients. User $C_1$'s application is compiled on the server with 100-cooperative deadline compilation vs. with selfish deadline compilation. 
    }
    \label{fig:deadline-comp-execution time}
\end{figure}

Fig. \ref{fig:deadline-comp-execution time} shows that varying user $C_1$'s BQC application compilation strategy on the server does not result in a reduction in execution time. Deadlines have little impact for classical blocks since their execution time is already quite short ($<$1$\mu$s). For the quantum blocks, these deadlines are almost never used. In BQC before local quantum blocks can be executed, entanglement must first be established. Since entanglement generation can only occur during specific slots in the network schedule, the availability of quantum blocks to be executed is staggered, which causes their deadlines to remain unused. The other user's ($C_2$, ..., $C_5$) BQC applications execution time and success probability remain constant regardless of user $C_1$'s compilation strategy on the server. And the success probability for user $C_1$ is unaffected. Similar results are observed for 2, 3, and 4 total clients. Ultimately, deadlines are irrelevant in this situation, as entanglement generation and the network schedule strongly influence node scheduling.

The large error bars can be attributed to the network schedule. Consider user $C_{1}$'s application for $n=3$ and a network schedule with bin lengths of 3. Once the simulation begins, all applications begin executing immediately. If user $C_{1}$'s application is first in the network schedule, all entangled pairs will be generated and the application finishes shortly after. However, if user $C_{1}$'s application is last in the network schedule, it waits for $c-1$ time bins before generating entanglement and completing execution. Since entanglement generation is the dominating factor in execution time (roughly 88\% of execution time for $n=3$), permutations of the network schedule cause significant variation of execution time.

\begin{table*}[]
    \centering
    \caption{Critical section execution time impacts of compiling user $C_{1}$'s BQC application with vs. without critical sections. 
    }
    \label{tbl:critical-comp:alt}
    \resizebox{0.75\textwidth}{!}{
        \begin{tabular}{|c|c|c|c|c|}
            \hline
            \multirow{2}{*}{\shortstack{\\[1pt] Number of \\ clients ($c$)}} & \multicolumn{2}{c|}{\rule{0pt}{2.5mm} BQC Program Size $n=3$} & \multicolumn{2}{c|}{\rule{0pt}{2.5mm} BQC Program Size $n=5$} \\
            \cline{2-5} 
             &  \begin{tabular}[c]{@{}c@{}} \rule{0pt}{2.5mm} $C_{1}$ Execution \\ Time Decrease \end{tabular}  
             &  \begin{tabular}[c]{@{}c@{}} \rule{0pt}{2.5mm} Avg $C_{i}$ Execution \\ Time Increase \end{tabular}  
             &  \begin{tabular}[c]{@{}c@{}} \rule{0pt}{2.5mm}  $C_{1}$ Execution \\ Time Decrease \end{tabular} 
             & \begin{tabular}[c]{@{}c@{}} \rule{0pt}{2.5mm}  Avg $C_{i}$ Execution \\ Time Increase \end{tabular}  \\ 
            \hline
             2 & \pct{5.05}{3.3} & \pct{18.90}{11.83}  & \pct{5.85}{1.40} & \pct{20.58}{3.10} \\ 
            \hline
              3 & \pct{5.57}{2.81} & \pct{16.84}{11.65} & \pct{6.12}{1.66} & \pct{18.11}{2.79} \\ 
            \hline
              4 & \pct{5.17}{3.31} & \pct{16.92}{11.29} & \pct{4.77}{1.88} & \pct{17.23}{2.23} \\ 
            \hline
              5 & \pct{4.38}{2.87} & \pct{15.44}{11.10} & \pct{4.99}{1.59} & \pct{16.51}{3.68} \\ 
            \hline
             
        \end{tabular}
    }
\end{table*}

\section{Critical Section Compilation}
\label{sec:critical-comp}
We consider an identical setup to deadline compilation (see \ref{sec:deadline-comp}), except now the default compilation strategy for low priority users is to not use critical sections (see \ref{sec:background:defs:crit}). The compilation strategy for high priority users is to compile with critical sections. Using critical section compilation for user $C_1$'s server program enforces that after entanglement generation is completed, the remaining blocks (the last four blocks for the optimized server program in Fig. \ref{fig:hybrid-bqc-ex}) will be executed without interleaving. When all of the server programs are compiled without critical sections we expect no difference in execution time or success probability between $C_1$ and the other users. However, when $C_1$ is a high priority user compiling with critical sections, we expect they will experience a decrease in execution time.

\subsubsection{Simulation Configuration}Our configuration is identical to that of Block Compilation Scenario 1 (see \ref{sec:block-comp:scen1:sim-config}).

\subsubsection{Results}
\label{sec:critical-comp:results}

\begin{figure}
    \centering
    \includegraphics[width=\linewidth]{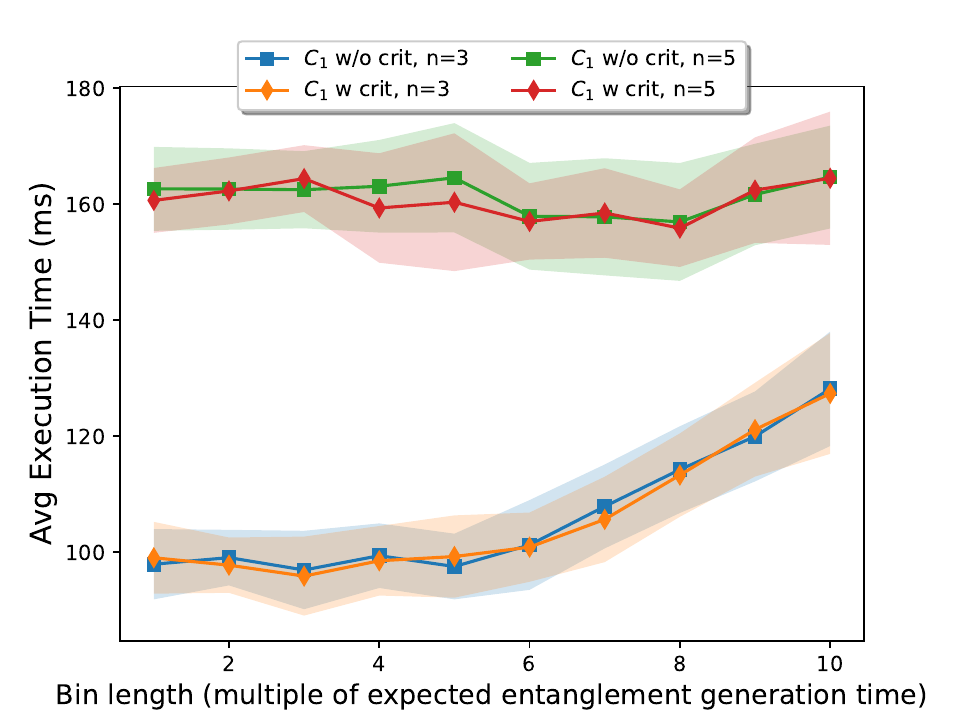}
    \caption{Critical section compilation execution time results for user $C_1$'s BQC application with a total of 2 clients. User $C_1$'s application is compiled on the server with vs. without critical section compilation. 
    }
    \label{fig:critical-comp-execution time}
\end{figure}

\begin{figure}
    \centering
    \includegraphics[width=\linewidth]{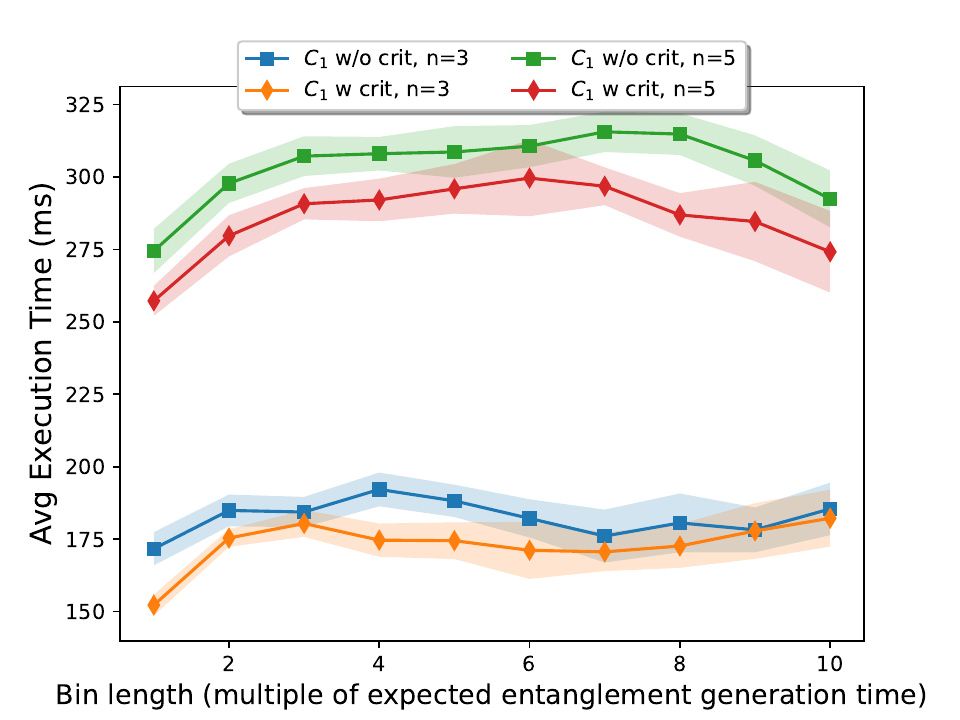}
    \caption{Critical section compilation execution time results for user $C_1$'s BQC application with a total of 2 clients. Local quantum blocks are modified to have a duration of 12-18\% of expected entanglement generation time (0.33-0.5\% previously). User $C_1$'s application is compiled on the server with vs. without critical section compilation. 
    }
    \label{fig:critical-comp-execution time-alt}
\end{figure}

Fig. \ref{fig:critical-comp-execution time} shows no advantage for user $C_{1}$ when using critical section compilation. Once entanglement has been generated, the blocks until the final measurement are short (0.33-0.5\% of expected entanglement generation time) and easily interleaved with the blocks of other programs. However, critical sections cause the other server program to have an increased execution time (9.93\% $n=3$, 9.57\% $n=5$, 2 clients). By preventing interleaving, all other BQC programs on the server are forced to halt their execution until $C_1$'s last measurement. We observe similar results for 3, 4, and 5 clients. 

We hypothesize that quantum programs with larger quantum blocks (i.e. more quantum gates) would see a benefit from critical sections due to reduced interleaving. We increase size of quantum blocks to vary between 12-18\% of expected entanglement generation time (previously 0.33-0.5\%) and re-run the simulations, showing the results in Fig. \ref{fig:critical-comp-execution time-alt}. When compiled with critical sections, user $C_1$'s BQC application experiences a decrease in execution time of up to 6.12\% (see Table \ref{tbl:critical-comp:alt}). Once again, the other BQC applications on the server experience an increase in execution time. The cause of the large errors bars is the same as \ref{sec:deadline-comp:results}.

\section{Conclusion}
\label{sec:conclusion}
Our work opens the door to compiler optimizations for quantum network programs.
Further work could include expanding on the present compilation ideas, or developing entirely new ones. It would also be interesting to understand how such techniques perform on other existing applications.
Additional hardware constraints, such as limited quantum memories and restrictions on which qubits can be used to generate entanglement also provide avenues for further research.

\section{Acknowledgments}
\label{sec:acknowledgements}
This work was funded by the European Union's Horizon Europe research and innovation programme under grant agreement No. 101102140 – QIA Phase 1.
We also acknowledge funding from the NWO VICI grant.
We thank Sacha Bernheim, Luca Marchese, and Diego Rivera for valuable feedback on an earlier version of this manuscript.

\bibliographystyle{unsrt}
\bibliography{references.bib}

\appendix
\section{Appendix}
\label{sec:appendix}

\subsection{Parameter Description}
\label{sec:appendix:params}

The Qoala simulator \cite{qoalasim} allows for low level details when configuring experiments. Below we describe each configurable parameter and the default values we chose for our experiments. 
\begin{itemize}
    \item \textbf{Dephasing time ($T_2$)} is a measure of the amount of dephasing noise a qubit in memory experiences. This value varies for different hardware platforms. We use a $T_2$ time of 10s for our simulations \cite{private}.
    
    \item \textbf{Gate duration} is the amount of time that it takes to execute a quantum gate (excluding the instruction execution time). These values can be unique for different hardware setups and for different types of gates (e.g. $X$ gate vs. $H$ gate). In our work we make the simplifying assumption that gates operating on the same number of qubits have identical gate durations. For our trapped ion hardware configuration we set the single qubit gate duration to 26.6$\mu$s \cite{avis_2023_reqprocnoderepeater} and the two qubit gate duration to 107$\mu$s \cite{krutyanskiy_2023_trappedionrepeater}.
    
    \item \textbf{Quantum instruction execution time} is the amount of time that it takes the quantum processing unit (QPU) to execute an instruction (excluding the gate duration). In the QNodeOS experiment \cite{delle2025operating}, the average instruction execution time was 50$\mu$s.
    
    \item \textbf{Gate fidelity} measures how accurately a quantum gate performs its intended operation compared to the ideal operation. The values for gate fidelity can be unique for different hardware platforms and for different gates. In our work we make the simplifying assumption that gates operating on the same number of qubits have identical gate fidelities. For our trapped ion hardware configuration we set the single qubit gate fidelity to 0.99 \cite{avis_2023_reqprocnoderepeater} and the two qubit gate fidelity to 0.95 \cite{krutyanskiy_2023_trappedionrepeater}.

    \item \textbf{Network Schedule} specifies sequential time slots when each application being executed in a quantum network is allowed to generate entanglement. These time slots are referred to as \emph{time bins}, and the duration of each slot is referred to as the \emph{bin length}. 
    In this work we assume that the network schedule follows a fixed repeating pattern. For example when considering a network schedule for three applications, ($A$,$B$,$C$), a pattern may be $bin_A$, $bin_B$, $bin_C$, $bin_A$, $bin_B$, \dots.
    
    \item \textbf{Bin Length} is how long a time bin in the network schedule lasts. In this work we assume all time bins have identical lengths. We use bin lengths that are an integer multiple of the expected time to generate entanglement.
    
    \item \textbf{Entanglement generation} is modeled as stochastic process in our simulator. Our process follows a geometric distribution, where each entanglement generation attempt succeeds with some probability, $p_{succ}$ and each attempt takes a fixed amount of time $t_{cycle}$. These vary for different hardware setups. When attempting to generate an entangled pair, a node will repeatedly execute entanglement generation attempts until an attempt succeeds, or the end of the time bin is reached.
    
    Below we describe the model and values used in our trapped-ion hardware configuration \cite{private}.

    \begin{multline}
        p_{succ} = \frac{1}{2} \eta_{penalty} \left(\eta_{ion} \eta^{ion\rightarrow telecom}_{FC} \eta^{telecom}_{det}\right)^2 
        \\ * 10^{-(\alpha / 10)(d/2)}
    \end{multline}

    \begin{equation}
        t_{cycle} = t_{class} + t^{ion}_{prep}
    \end{equation}
    
    \begin{itemize}
        \item $\alpha$ = 0.2: Attenuation factor at telecom frequency.
        \item $\eta_{ion}$: The efficiency for emitting and collecting a photon. Anticipated value: 0.5 / 0.87.
        \item $\eta^{ion\rightarrow telecom}_{FC}$: Frequency conversion efficiency from ion to telecom. Anticipated value: 0.7.
        \item $\eta^{telecom}_{det}$: Detection efficiency for photons at telecom frequency. Anticipated value: 0.9.
        \item $\eta_{penalty}$: A penalty paid to truncate the detection window, increasing the fidelity. Anticipated value: 0.2 with fidelity 0.95.
        \item $d$: The distance (km) between nodes.
        \item $t_{cycle}$: The average time per generation attempt.
        \item $t_{class}$: Classical communication time.
        \item $t^{ion}_{prep}$: Time to prepare the ion. Anticipated value 200$\mu s$.
    \end{itemize}
    The expected time to generate an entangled pair is $t_{cycle}/p_{succ}$.

    \item \textbf{Classical instruction execution time} is how long it takes the classical processing unit (CPS) to execute an instruction. In the QNodeOS experiment a desktop was used for the CPS (4 Intel 3.20GHz cores, 32 GB RAM, Ubuntu 18.04) \cite{delle2025operating}. Each clock cycle takes 0.3125ns. Looking at the instruction set for an Intel Pentium, the most expensive single instruction (in terms of clock cycles) took 46 clock cycles \cite{fog_2022_instructiontables}. This would take roughly 15ns to execute on a single core of the CPS used in the QNodeOS experiment.

    \item \textbf{Node scheduling messaging processing time} is the communication time between the CPS/QPS schedulers and the node scheduler. As envisioned in the Qoala extension \cite{vandervecht2025qoala}, we expect this to be implemented with shared memory. We use a value of 60ns derived from shared memory access taking 192 clock cycles \cite{maroun_2019_sharedmem}. 

    \item \textbf{Classical communication latency} is the time it takes to send a classical message from one node to another over the network (e.g. via TCP). We calculate the classical communication latency (CCL) as a function of distance ($d$) and hop count ($h$) using the following formula \cite{bovy_2002_internetdelay, bozkurt_2018_fiberlatency}.
    \[\text{CCL} = h \cdot 244\mu s + 155\mu s + \frac{d}{200,000 km/s}\]
    
    \item \textbf{Classical communication processing time} is the additional time that it takes the CPS to receive a classical message. This requires an additional 478 clock cycles for a nonblocking sysrecvfrom call \cite{hruby_2014_socketsyscall}. This would take the CPS used in the QNodeOS experiment roughly 150ns.
    
    \begin{figure}
        \centering
        \includegraphics[width=0.7\linewidth]{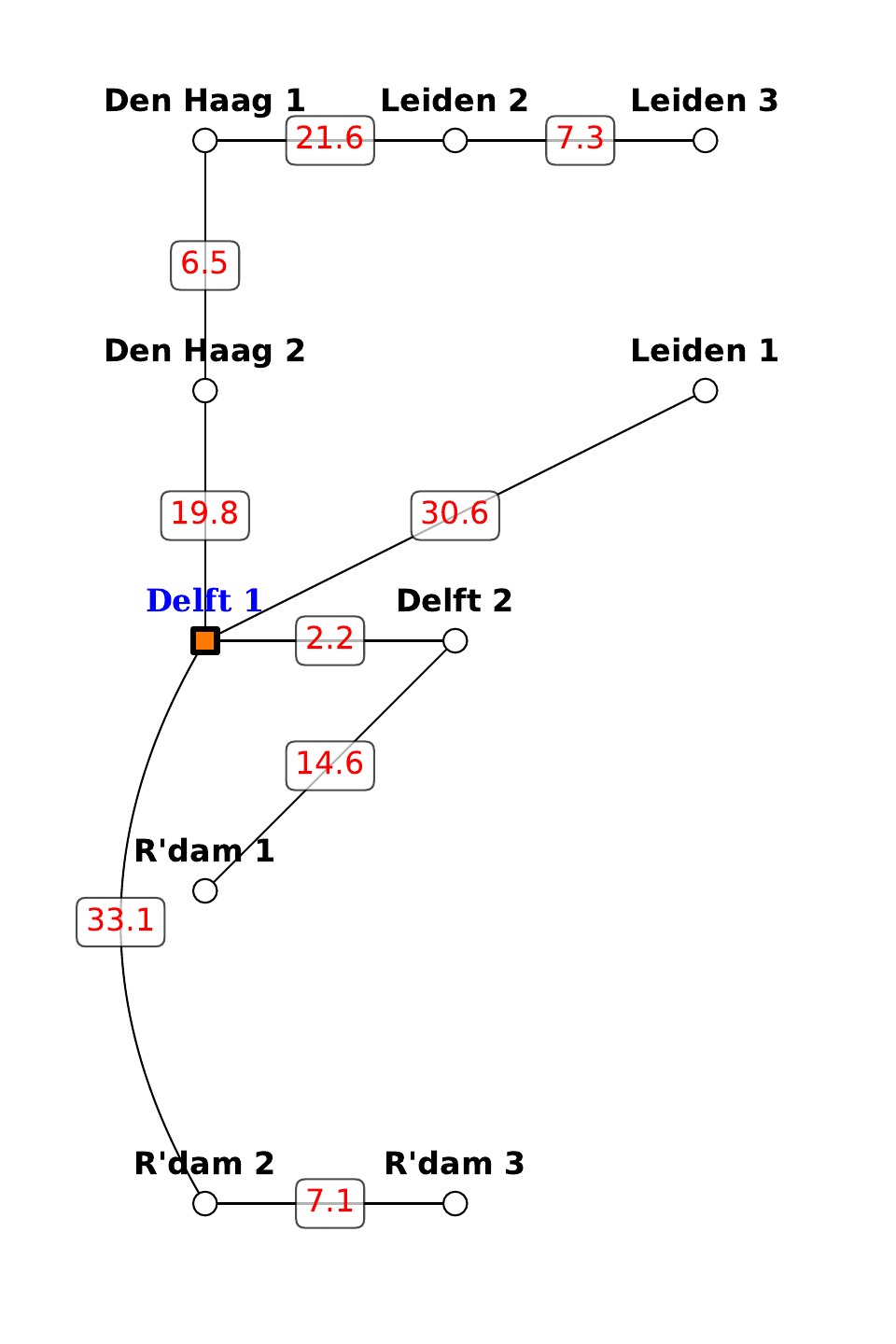}
        \caption{A partial topology of the SURF network \cite{rabbie_2020_surfnetrepo, rabbie_2022_surfnetinfra}. SURFnet is an education and research network provider in The Netherlands. The distances between nodes (km) are shown in red. We highlight the location of our server (Delft 1) in blue and vary the location of the client between the remaining nodes.}
        \label{fig:surfnet}
    \end{figure}
    
    \item \textbf{SURFnet} is an educational and research network provider in The Netherlands. We use the SURF network topology  (see Fig. \ref{fig:surfnet}) for the distance and number of hops between nodes \cite{rabbie_2020_surfnetrepo, rabbie_2022_surfnetinfra}. For entanglement generation, we assume that the nodes are directly connected, with the heralding stations centered directly at the midpoint of the link. For classical communication, we assume that each intermediate node adds a hop, resulting in additional classical communication delay. 
\end{itemize}

\end{document}